%% file: main.tex
\documentclass[sigconf,authorversion=true]{acmart}

\AtBeginDocument{%
  }

\usepackage{multirow}
\usepackage{booktabs}
\usepackage{tabularx}
\usepackage{caption}
\usepackage{subcaption}
\usepackage{balance}

\copyrightyear{2025}
\acmYear{2025}
\setcopyright{acmlicensed}
\acmConference[SIGIR '25]{Proceedings of the 48th International ACM SIGIR Conference on Research and Development in Information Retrieval}{July 13--18, 2025}{Padua, Italy} \acmBooktitle{Proceedings of the 48th International ACM SIGIR Conference on Research and Development in Information Retrieval (SIGIR '25), July 13--18, 2025, Padua, Italy} \acmDOI{10.1145/3726302.3730293}
\acmISBN{979-8-4007-1592-1/2025/07}

\acmConference[SIGIR '25]{Proceedings of the 48th International ACM SIGIR Conference on Research and Development in Information Retrieval}{July 13--18, 2025}{Padua, Italy}

\begin{document}

\title{Revisiting Algorithmic Audits of TikTok: Poor Reproducibility and Short-term Validity of Findings}


%

\author{Matej Mosnar}
\authornote{Both authors contributed equally to this research.}
\affiliation{%
  \institution{\small Kempelen Institute of Intelligent Technologies \normalsize}
  \city{Bratislava}
  \country{Slovakia}}
\email{matej.mosnar@kinit.sk}
\orcid{0009-0009-2466-7225}

\author{Adam Skurla}
\authornotemark[1]
\affiliation{%
  \institution{\small Kempelen Institute of Intelligent Technologies \normalsize}
  \city{Bratislava}
  \country{Slovakia}}
\email{adam.skurla@kinit.sk}
\orcid{0009-0000-9337-2804}

\author{Branislav Pecher}
\affiliation{%
  \institution{\small Kempelen Institute of Intelligent Technologies \normalsize}
  \city{Bratislava}
  \country{Slovakia}}
\email{branislav.pecher@kinit.sk}
\orcid{0000-0003-0344-8620}

\author{Matus Tibensky}
\affiliation{%
  \institution{\small Kempelen Institute of Intelligent Technologies \normalsize}
  \city{Bratislava}
  \country{Slovakia}}
\email{matus.tibensky@kinit.sk}
\orcid{0000-0001-9928-3474}

\author{Jan Jakubcik}
\affiliation{%
  \institution{\small Kempelen Institute of Intelligent Technologies \normalsize}
  \city{Bratislava}
  \country{Slovakia}}
\email{jan.jakubcik@kinit.sk}
\orcid{0009-0009-5162-2858}

\author{Adrian Bindas}
\affiliation{%
  \institution{\small Kempelen Institute of Intelligent Technologies \normalsize}
  \city{Bratislava}
  \country{Slovakia}}
\email{adrian.bindas@kinit.sk}
\orcid{0009-0002-8796-6475}

\author{Peter Sakalik}
\affiliation{%
  \institution{\small Kempelen Institute of Intelligent Technologies \normalsize}
  \city{Bratislava}
  \country{Slovakia}}
\email{peter.sakalik@kinit.sk}
\orcid{0009-0005-8959-3838}

\author{Ivan Srba}
\affiliation{%
  \institution{\small Kempelen Institute of Intelligent Technologies \normalsize}
  \city{Bratislava}
  \country{Slovakia}}
\email{ivan.srba@kinit.sk}
\orcid{0000-0003-3511-5337}

\renewcommand{\shortauthors}{Mosnar et al.}

\begin{abstract}
  Social media platforms are constantly shifting towards algorithmically curated content based on implicit or explicit user feedback. Regulators, as well as researchers, are calling for systematic social media algorithmic audits as this shift leads to enclosing users in filter bubbles and leading them to more problematic content. An important aspect of such audits is the reproducibility and generalisability of their findings, as it allows to draw verifiable conclusions and audit potential changes in algorithms over time. In this work, we study the reproducibility of the existing sockpuppeting audits of TikTok recommender systems, and the generalizability of their findings. In our efforts to reproduce the previous works, we find multiple challenges stemming from social media platform changes and content evolution, but also the research works themselves. These drawbacks limit the audit reproducibility and require an extensive effort altogether with inevitable adjustments to the auditing methodology. Our experiments also reveal that these one-shot audit findings often hold only in the short term, implying that the reproducibility and generalizability of the audits heavily depend on the methodological choices and the state of algorithms and content on the platform. This highlights the importance of reproducible audits that allow us to determine how the situation changes in time.
\end{abstract}




\maketitle


\input{1-introduction}
\input{2-related_work}

\input{3-methodology}
\input{4-results}

\input{5-ethical-considerations}
\input{6-conclusion}




\begin{acks}
This work was partially funded by the EU NextGenerationEU through the Recovery and Resilience Plan for Slovakia under the project \textit{AI-Auditology}, No. 09I03-03-V03-00020; and by the European Union under the Horizon Europe project \textit{vera.ai}, GA No. \href{https://doi.org/10.3030/101070093}{101070093}.

In addition, we would like to thank Marek Havrila, Simon Liska, Andrej Suty and Filip Hossner for their help in preparing the necessary codebase for running the experiments and Matus Mesarcik, Katarina Marcincinova and Juraj Podrouzek for their help with ethical and legal considerations of this work.
\end{acks}

\bibliographystyle{ACM-Reference-Format}
\balance
\bibliography{references}

\appendix

\end{document}

%% file: 1-introduction.tex
\section{Introduction}
\label{sec:introduction}

Social media platforms are pervasive in our lives, with their popularity and the amount of user-generated content on them rapidly increasing. In recent years, these platforms have shifted from simply providing chronologically ordered content based on users' connection towards algorithmically curated content that does not require extensive and explicit user feedback~\cite{klug2021trickplease}. At the same time, the content on the platform has shifted to a short-video format (e.g., videos that are 30-60 seconds long). Together, these changes lead to a shift in how users consume and engage with social media content, e.g., increasing the time spent on the platform~\cite{zannettou2024analyzing}. However, such a shift leads to significant problems as well. First, the algorithmically curated recommendations are viewed as non-transparent to end users. In addition, they tend to enclose users in filter bubbles, showing only the content that achieves strong engagement, which leads to a proliferation of highly problematic content such as misinformation or extreme and polarising viewpoints~\cite{Ribeiro2020, tomlein_audit_2021}.

To address these problems, researchers started performing social media algorithmic auditing studies~\cite{urman2024mapping, bandy2021survey, sandvig2014auditing} to better understand the algorithms and ways to mitigate the potential negative consequences. Significant focus was dedicated to the understanding of the "traditional" social media and their content, such as Facebook, YouTube as well as different search engines~\cite{boeker2022empirical, vombatkere2024tiktok, mousavi2024auditing, Spinelli2020, Hussein2020, Papadamou2020, tomlein_audit_2021}. These studies have confirmed the negative impact of social media, where the algorithm unintentionally leads users to more and more problematic content, as it generates high levels of engagement. However, the auditing studies for the short-video format platforms (e.g., TikTok) are only starting to slowly appear. 

At the same time, to mitigate the negative consequences, new policies are enacted -- most prominently the Digital Services Act (DSA)~\cite{DSA}. This legislation calls for systematic audits of social media platforms (according to Article 37) to assess the obligations that social media platforms need to follow -- some of them are specifically dedicated to algorithms that curate content on these platforms. Algorithmic audits represent one of the means by which algorithmic-related obligations of social media platforms can be assessed. An important aspect of such algorithmic audits is their reproducibility and generalisability of their results, even as the content on the platform may evolve and change. Without reproducibility, the findings of the algorithmic audits are not actionable as they do not allow the policymakers and regulators to demand and consequently verify changes on the platforms.

\textbf{The main goal of this study is to determine the reproducibility of the social media algorithmic audits and the generalisability of their results when performed after some period of time and in extended settings (i.e., in additional countries).} For this purpose, we reproduce and extend the previous works that audit the popular, short-video format social media platform TikTok. We chose TikTok due to its popularity (mainly among younger audiences) and the strength and complexity of its recommender (i.e., it can quickly determine user interests and steer users towards niche content). Specifically, we build on the works of \citet{boeker2022empirical}, \citet{vombatkere2024tiktok} and partially also \citet{mousavi2024auditing} (in terms of evaluation metrics). The primary aim of these reference studies was to investigate how \textit{personalisation factors}, such as user location or various user interactions with content, influence the level of personalisation or explanations of why the videos were recommended. All of them audit the recommender system behind the \textit{For You} section of the TikTok platform.

Since the previous studies were performed before the enactment of the DSA, the policy may already have an impact on the platforms. At the same time, the recommender systems, as well as the content itself, evolve constantly. As all of these factors may affect the nature of recommended items, it is important to continuously reproduce the findings (or use a longitudinal algorithmic audit study).

By revisiting the experiments from the reference studies, we address the following research questions:
\begin{itemize}
    \item \textbf{(RQ1)} What is the \textit{level of reproducibility} of algorithmic audits taking into consideration the constantly evolving nature of social media platforms?
    \item \textbf{(RQ2)} What is the \textit{effect of different personalisation factors}, such as location, watch duration, liking and following, on the \textit{For You} video recommendation on TikTok? How has the importance of different personalisation factors \textit{evolved since the reference studies}?
\end{itemize}

To answer these research questions, while trying to reproduce the reference studies as faithfully as possible, we also did several extensions and some inevitable deviations, specifically: 1) extending the setting to different countries, as the behaviour of the recommender system may differ across them~\cite{urman2024mapping}; 2) adjusted and fixed specific methodological decisions (e.g., watch time being lower for the videos of interest, or dealing with shadow-bans) that may have unintentionally affected the findings~\cite{chandio2024audit}; 3) updated the algorithmic audit implementation based on the platform changes (e.g., dealing with ads and live streams, or dealing with the changes in the underlying HTML structure) in order to make the audit even possible; and 4) providing additional analysis of the collected data in order to draw additional findings and allow for better understanding of the platform behaviour, including a post-audit check using bot interaction data obtained through GDPR requests.

Overall, our main contributions and findings are as follows:
\begin{itemize}
    \item We reproduce the previous sockpuppeting studies that investigate the personalisation factors on TikTok's \textit{For You} page. We identify multiple factors that severely and negatively affect the reproducibility and generalisability of the algorithmic audits -- causing us to spend over the period of 5 months roughly 9 person months on replication. As such, we highlight the need for a change in the overall paradigm of how audits are performed -- ad-hoc, without any established methodology and in a single-shot manner.
    
    \item By reimplementing and rerunning the algorithmic audit after 3 years, we find that the situation has changed significantly since the reference studies. We observe that the \textit{watch} action provides the strongest personalisation impact, which is increased as the video is watched longer or multiple times. At the same time, we observe a stronger exploration aspect for the explicit actions (\textit{like} and \textit{follow}) at the start, followed by a strong exploitation phase. However, post-audit analysis of GDPR data, revealed that explicit actions may not be taken into consideration by the recommender in some audit scenarios, which may bias these results. At the same time, follow action does not seem to work on the web interface even for non-bot accounts. Furthermore, we observe that the findings strongly depend on the methodological setup. Just by choosing a different evaluation metric, we can observe completely different findings. A similar effect can be observed based on small variations of the user's simulation, such as choosing slightly different topics of interest or what kind of actions and how often they are performed.  
    
    \item To allow for future reproducibility, we publicly release the code of our audit, along with the data we have gathered and the analysis we performed\footnote{\url{https://github.com/kinit-sk/tiktok-algorithmic-audit-reproducibility}}.
\end{itemize}

Following the findings of this paper, we advocate for \textbf{reproducible, longitudinal, multiplatform and more authentic (in terms of user simulation) audits} that can more faithfully discern whether the change in findings is only a result of content change or due to enacted policies and algorithm changes. A way towards this kind of continuous audit is inevitably connected with their automatisation (especially in terms of content labelling) ~\cite{simko2021continuous-automatic-audits}.

%% file: 2-related_work.tex
\section{Related Work}
\label{sec:related-work}

A large number of works perform algorithmic audits of the social media recommender systems~\cite{urman2024mapping, bandy2021survey, sandvig2014auditing}. The focus of these works is mainly on the distortion and disparities in content delivery that arise due to the personalisation of the recommender systems~\cite{urman2024mapping}. A popular direction is to better understand the effects of different personalisation factors, such as gender, age, location, or the different interactions with the content, on the recommendation~\cite{boeker2022empirical, kaplan2024comprehensively, vombatkere2024tiktok, robertson2018auditing, evans2023google, thorson2021algorithmic, Hussein2020}. Although the explicit feedback (liking and following) often has a larger impact, even just watching videos for a set period of time (implicit feedback) is a strong indicator for the personalisation~\cite{boeker2022empirical, vombatkere2024tiktok}. Multiple works focus on investigating the proliferation of harmful content and how the recommendations lead users to more extreme content~\cite{Spinelli2020, Ribeiro2020, juneja2023assessing, zieringer2023algorithmic, haroonYouTubeGreatRadicalizer2022, ballardConspiracyBrokersUnderstanding2022}. Previous studies have even found that following the chain of recommendations leads users away from reliable sources and towards extreme content~\cite{Spinelli2020, Ribeiro2020}. Similarly, multiple works focus on better understanding of how the recommender systems enclose users in filter bubbles and how such bubbles can be dealt with~\cite{tomlein_audit_2021, srba2023auditing, Hussein2020, yang2023bubbles, ledwich2022radical, kaiser2020birds, aridor2020deconstructing}. These studies found that even though the tendency to enclose users in filter bubbles is strong, the filter bubbles can be reversed by explicitly switching to another content~\cite{tomlein_audit_2021, srba2023auditing, yang2023bubbles}.

When collecting the data for analysis, researchers rely on different methodologies. The first type of so-called \textit{sockpuppeting audits} employs automated \textit{bots} to simulate the user behaviour on the platform~\cite{tomlein_audit_2021, boeker2022empirical, vombatkere2024tiktok, mousavi2024auditing}. Although it leads to better control over the algorithmic audit, the simulated behaviour is often very artificial and unrealistic, and the bots lack authenticity and diversity of real users~\cite{vombatkere2024tiktok}. Another possibility is to use the API of the platforms to collect the data on the platform~\cite{Papadamou2020, chuai2024changes, pinto2024tracking}. However, collecting data in such a way can only be used to analyse the content on the platform, but not the behaviour of the underlying algorithm. Finally, specific studies have recently started using donated traces from real-world users that can be obtained through a request under the GDPR~\cite{zannettou2024analyzing, vombatkere2024tiktok}. Although it represents the most natural behaviour of users on the platform, it does not allow for the controlled investigation of specific scenarios. These methodological decisions in how the data is collected significantly impact the overall accuracy and reliability of the audit findings~\cite{chandio2024audit}.

When it comes to the audited platforms, the majority of the focus is on search engines (Google, Bing) or YouTube, with only a limited focus on TikTok, the most popular short-term video format platform, while there are no audits on Instagram or YouTube Shorts~\cite{urman2024mapping}. \citet{kaplan2024comprehensively} analysed the role of demographics through the TikTok mobile app, finding that the content and ads on the platform are mostly affected by age. The authors encountered various challenges when performing the algorithmic audits through the app, such as the need to intercept the network traffic. \citet{boeker2022empirical} investigated the effects of different personalisation factors, such as language, watch duration, and like and follow actions, on the recommender system. The authors find that the follow action has the most impact, while a simple irreversible watch action has the same strong influence as liking the video. As such, users can be led into filter bubbles and harmful or radicalising content by simply lingering over a problematic video for some time~\cite{boeker2022empirical}. \citet{mousavi2024auditing} analysed the faithfulness of the explanations provided by TikTok on why the specific video was recommended. By analysing different engagement types (no engagement, liking, sharing, following), the authors find that in many cases the explanations are very generic ("video is popular in your country") or reference actions that were not performed ("commented on a similar video"). All of these works utilised automated sockpuppeting bots to perform the analyses. On the other hand,~\citet{zannettou2024analyzing} used a watch history of real TikTok users to analyse the changes in the behaviour of such users. They found that the majority of the videos are not watched in full, and the liking of recommended videos is increasing over time. Finally,~\citet{vombatkere2024tiktok} used both sockpuppeting and donated data to analyse the exploration and exploitation on the TikTok feed. The authors found that only around 30-50\% of the videos are personalised based on user interests and interactions while observing similar importance of individual personalisation factors as~\cite{boeker2022empirical}.

In this work, we focus on reproducing the works described in the previous paragraph, focusing mainly on the works of~\cite{boeker2022empirical, vombatkere2024tiktok} and partially also~\cite{mousavi2024auditing}. We report on the reproducibility struggles and the changes in observed behaviour and findings when running the algorithmic audits after a period of time (in January and February 2025, more than 3.5 years from the reference study \cite{boeker2022empirical}).

%% file: 3-methodology.tex
\section{Study Design and Methodology}
\label{sec:methodology}

Reproducibility is one of the most important aspects of any algorithmic audit. However, many factors may affect reproducibility, both from the perspective of the social media platform and the audits themselves, which may finally influence the reproducibility and generalisability of the findings. As a case study for studying the level of algorithmic audit reproducibility (and the factors that negatively affect it), we take mainly the works of~\citet{boeker2022empirical} and~\citet{vombatkere2024tiktok} (denoted onwards as the \textit{reference studies}). Authors of both works perform an agent-based sockpuppeting algorithmic audit of the personalisation factors on TikTok, which leads to similar findings. 

While reusing the methodology from reference studies (and, if possible, the audit code), we run into a number of reproducibility issues, some even preventing us from effectively performing an audit. For this reason, we developed an adjusted study design and methodology, and furthermore, we also addressed the drawbacks we identified in these studies (see Section~\ref{sec:results_rq1} for details). Specifically, we reimplemented the code of the audit to take into consideration the platform changes, modified the methodology for investigating the different personalisation factors to mitigate the possible issues and better explore the generalisability of the findings, and adjusted the bot interests to be more up-to-date. In addition, we also introduced an improved method for analysing the collected data. 

All these changes and extensive practical experience finally allowed us to answer the first research question (RQ1) by identifying and describing the factors that limited reproducibility during the whole process. Adjusted and improved methodology allowed us to conduct the audit itself, and as such, we were able to answer the second research question (RQ2) by comparing the effect of personalisation factors between our study and the reference studies.

Overall, to answer our research questions and investigate the importance of different personalisation factors on TikTok, we conducted an agent-based sockpuppeting audit study. We let a series of agents interact with the content recommended on the TikTok \textit{For You} page within the platform's web-based user interface. For each bot, we created a brand new user account with a random name, email address, date of birth and a predetermined location. 

To assess the importance of the personalisation factors, we defined various \textit{audit scenarios} that describe and modify the general behaviour and/or the characteristics of the bot. To this end, we reused scenarios defined in the reference study~\cite{boeker2022empirical}, while we purposefully omitted some redundant or overlapping scenarios to make reproducibility more efficient. Specifically, we run the following scenarios that can be mapped to individual personalisation factors, which have shown a significant impact in the reference studies: 1) none (to control for noise); 2) location; 3) watch duration; 4) liking; and 5) following. In each scenario, we run two separate bots at the same time, one serving as a \textit{control} user (called \textit{control} from now on) and the other one that is used to determine the impact of personalisation factors (called \textit{personalised} from now on). Each scenario consists of 4 runs (sessions) while keeping the history of the users (i.e., for the specific scenario, the same user logs in and interacts with a predetermined number of videos 4 times, with a longer break between each run -- approx. one day-long). This break allows the TikTok recommender system to reflect on the previous bots' actions and develop personalisation. To keep the history, we both reuse the cookies (saved at the end of the previous run) or login to the user again using the platform interface. The data collection is performed across January-February 2025. Finally, some selected scenarios have been repeated two times with a different set of control and personalised users. In this way, we verified that the same conclusions can be drawn when the audit is repeated in a short period of time, and an inherent randomness of the recommendation system would not invalidate findings. The results have shown no significant differences between these repeats, thus proving the robustness of our methodology and implementation.

The full scenario list, with all the details, is available in Table~\ref{tab:bot-scenarios}. Below, we describe the individual setup for the scenarios.

\newcolumntype{Y}{>{\centering\arraybackslash}p}
\begin{table*}[]
\caption{Scenarios for investigating the effects of personalisation factors, with information on how long the bot watches the videos and how many times the scenario is repeated (column Rep.). Note that the "control" specifies both the watchtime of the control user as well as the videos deemed "not relevant" for the personalised user. [S\#] denotes ID of each scenario.}
\label{tab:bot-scenarios}
\begin{tabularx}{\textwidth}{@{}p{0.06\textwidth}Xp{0.02\textwidth}Y{0.15\textwidth}p{0.03\textwidth}@{}}

\toprule

Factor                    & Scenario setup       & ID  & \begin{tabular}[x]{@{}c@{}}Watchtime \\ Control | Personalised \end{tabular}                       & Rep. \\ \midrule

\textbf{None}                   & Country: USA                   & [S0] & 100\% | N/A                                       & 1                       \\ \midrule
\multirow[c]{5}{*}{\textbf{Location}} & Country: Germany               & [S1] & 100\% | 100\%              & 1                  \\
                          & Country: France                & [S2] 
 & 100\% | 100\%                                          & 1 \\
                          & Country: Romania               & [S3] & 100\% | 100\%                                          & 1 \\
                          & Country: Ukraine               & [S4] & 100\% | 100\%                                          & 1 \\ \midrule
\multirow[c]{7}{*}{\textbf{Watch}}    & 25 random videos             & [S5] & 25\% |  50\%                 & 1                       \\
                          & 25 random videos             & [S6] & 25\% |  75\%                                             & 1 \\
                          & 25 random videos             & [S7] & 25\% |  100\%                                            & 1 \\
                          & 25 random videos             & [S8] & 100\% |  200\%                                           & 1 \\
                          & \multirow{3}{\hsize}{Hashtag: movie, film, marvel, foodtiktok, tiktokfood, foodie, cooking, food, gaming, gta6, gta, minecraft, roblox, cat, dog, pet, dogsoftiktok, catsoftiktok, cute, puppy, dogs, cats, animals, petsoftiktok, kitten, comedy, lol, humour, laugh, fun, jokes, love, couple, relationships} & [S9] & 25\% |  50\%       & 1                       \\
                          &                                & [S10] & 100\% |  200\%                                         & 2 \\
                          &                                & [S11] & 100\% |  400\%                                         & 2 \\ \midrule
\multirow[c]{3}{*}{\textbf{Like}} & 10 random videos             & [S12] & 100\% | 100\%                    & 1 \\
                          & Hashtag: \textit{Same as watch}                               & [S13] & 100\% | 100\%           & 2          \\
                          & Creator: Top 30 creators on TikTok  & [S14] & 100\% | 100\%                & 2 \\ \midrule

\multirow[c]{2}{*}{\textbf{Follow}}   & 5 random creators              & [S15] & 100\% | 100\%              & 1       \\
                          & 10 random creators             & [S16] & 100\% | 100\%                                          & 1  \\

\bottomrule
\end{tabularx}
\end{table*}

\textbf{Default scenario.} Default scenario [S0] describes the common setup (configuration) that is afterwards adapted for each more specific scenario. First, we initialise a new incognito Chrome browser session for each bot run, with the language set to English and the location set to the USA (using a proxy server). For interacting with the webpage, we use the nodriver\footnote{\url{https://github.com/ultrafunkamsterdam/nodriver}} python library. After initialising the session, the bot navigates to the TikTok website, logs in as the specific user, and handles any banners that come up (e.g., accepting cookies or banners prompting the bot to use a mobile phone application). Afterwards, the bot navigates to the \textit{For You} page and scrolls through 250 videos, watching each video for 100\% of its duration or up to a specific limit of 120 seconds, whichever is shorter. We set the watch time limit, as we are interested in the short videos, but have noticed in a pilot study that a small fraction of TikTok videos may be longer (even 1 hour or longer), or we get livestreams as part of the recommendations. This way, we guarantee that such videos are "skipped". At each video, we collect all available video metadata (title, description, hashtags, name of the video creator, number of watches, likes, comments or interactions, etc.). After watching the 250 videos, we terminate the session. In this scenario, the control user performs the exact same actions and has the same characteristics as the personalised user.

\textbf{Controlling for noise.} By executing the default scenario without any further modification, we measure the standard level of noise, i.e., we determine the typical rate of encountering unique videos across separate runs of the bots and between the control and personalised users (i.e.,  the percentage of videos that are expected not to be shared by the \textit{control} and \textit{personalised} users).

\textbf{Personalisation factor: Location.} When investigating the \textit{location} personalisation factor, we change the location of the personalised user to a different country using a proxy server. We choose the following countries: 1) Germany [S1]; 2) France [S2]; 3) Romania [S3]; and 4) Ukraine [S4]. The countries are chosen based on the proxy server availability and whether specific important events are happening in the countries (e.g., elections in Germany and Romania, war in Ukraine). The remaining setup stays the same as in the default scenario, i.e., watching each video for 100\% of its duration (or 120 seconds). The control user is located in the USA.

\textbf{Personalisation factor: Watch Duration.} As we consider the \textit{watch duration} personalisation factor to be the most important one (it requires no active user participation, only implicit feedback), we focus on a more comprehensive evaluation and define multiple possibilities of how we choose what videos will be watched longer and how long the bot will watch the video.

In the first set of scenarios, we choose the videos randomly. For each run, we randomly select 10\% of the videos (amounting to 25) that will be watched longer. Using this set of scenarios, we determine the overall importance of the watch duration personalisation, with no further systematic feedback (e.g., no specific user interest). To properly investigate the impact, the personalised users watch the selected videos for: 1) 50\% [S5]; 2) 75\% [S6]; 3) 100\% [S7]; or 4) 200\% [S8] of their overall length. The remaining videos are watched for 25\% (for [S5]-[S7]) or 100\% (for [S8]) of the overall duration. 

In the second set of scenarios, we choose the videos based on user interests. We follow the setup by~\citet{boeker2022empirical}, using a set of hashtags that define the user interests. The videos that match at least one of the hashtags are deemed "of interest" for the personalised user and watched for longer. In our study, we adjust the list of hashtags from~\citet{boeker2022empirical} to be more up-to-date (the hashtags are listed in Table~\ref{tab:bot-scenarios}). We opted for a set of hashtags that is more general and represents broader interests. In this case, we watch the videos for: 1) 50\% [S9]; 2) 200\% [S10]; or 4) 400\% [S11] of their duration. The remaining videos are watched for 25\% (for [S9]) or 100\% (for [S10] and [S11]) of their overall duration.

For the control user, we set the watch duration the same way as the "videos not of interest" in the personalised users -- watching for 25\% of the overall duration if the maximum watch time of the personalised user is less or equal to 100\%, or 100\% otherwise.

\textbf{Personalisation factor: Liking.} When investigating the \textit{like} personalisation factor, we explore 3 separate scenarios for choosing the videos the personalised user will like. First, the bot randomly chooses 10 videos in each run [S12]. This case is similar to the randomised video watch, providing no further systematic feedback. Second, we choose the videos based on the user's interests [S13]. The user interests are defined the same way as in the video watch scenario -- using the same list of hashtags. As such, the bot will like any video that matches at least one of the hashtags. Third, the bot likes all the videos that are uploaded by a specific creator [S14]. We chose the 30 most popular creators on the platform at the time of the experiment in order to maximise the probability of encountering such videos. The control user just watches each video following the default scenario setup. As such, the only difference between control and personalised user is the like action.

\textbf{Personalisation factor: Following.} For the \textit{follow} personalisation factor, we introduce a follow action for the personalised user. In each run, the personalised user follows 5 [S15] or 10 [S16] random creators from the videos that are encountered. The remaining setup for the personalised and control user stays the same.

\textbf{Evaluating the collected data.}  Following the reference studies, we utilise only heuristics instead of annotations. Although annotating the data would lead to more precise results, it requires significantly larger effort and limits the study's extent. Also, sticking to the same evaluation approach allows us to compare findings with the reference studies directly. We mainly rely on two video characteristics to analyse the personalisation from different factors.

First, we analyse the \textbf{video popularity} represented by the video play count. Previous studies have observed (e.g., ~\cite{klug2021trickplease}) that with time, due to personalisation, the play count of the videos the users observe decreases as the algorithm starts showing more niche content. As such, we can measure the personalisation effect as the decrease in the average video popularity, where a larger decrease means stronger personalisation.

Second, the \textbf{video hashtags} are a simple way to represent the video content. As such, we utilise the Jaccard similarity between videos based on their hashtags as another metric. We expect that as the TikTok algorithm serves more personalised videos, the similarity between these videos will start increasing (as they will deal with similar content). To calculate the similarity using hashtags, we first preprocess all of them, by making them lowercase, removing non-alphanumerical symbols and removing very generic hashtags (such as "fyp" or "foryoupage"; the list is available as part of the released codebase). In addition, we define similarity measures with different strictness levels to provide further analysis. First, we use the original \textit{Jaccard similarity} index that measures a ratio of similar hashtags across a pair of videos. For example, if videos share 2 hashtags out of 10, the similarity is measured as 20\%. Second, a more lenient measure we use is whether the videos have at least one common hashtag (as used by~\cite{mousavi2024auditing}). We further modify this measure by including substrings in the search, as we have observed that many videos may use longer hashtags occasionally even though they represent the same topic (e.g., "cat" and "cutecat"). As such, if at least one hashtag matches or a hashtag in one video is a substring of a hashtag in the paired video, we consider the videos to be similar and assign 100\% similarity. We call this the \textit{basic match similarity}.

To provide our analysis, we use both hashtag similarity measures to calculate the similarity between sets of multiple videos. For example, we compare the similarity between the video feeds of the control user and the personalised user (expecting them to diverge more and more) or the similarity between videos at the start of the scenario and the end (expecting the videos of the control user to be similar to each other throughout the whole scenario and the videos of the personalised user becoming more dissimilar to the start where no personalisation should be present). To calculate this similarity, we repeat the above-mentioned calculation for each pair of videos from the two sets and calculate an average similarity.

%% file: 4-results.tex
\section{RQ1: Reproducibility of Algorithmic Audits}
\label{sec:results_rq1}

Following our effort to reproduce the references studies, we identified two sets of factors negatively affecting audit reproducibility: 1) \textit{audit-specific} that stem from the audit itself and its methodology; and 2) \textit{platform-specific} that are caused by the platform itself.

\textbf{Audit-specific.} The most important factor limiting the reproducibility is the availability and quality of the resources that were used to run the audit -- the code used for running the agent-based sockpuppeting bots and the videos used or encountered during the audit. Even though the reproducibility of audits is paramount, the \textbf{audit works often do not publicly release these resources or the published source code is incomplete and missing fundamental parts.} In this study, we had to reverse-engineer specific details from the source code (such as a database schema description) that were completely missing in the code published in~\cite{boeker2022empirical}. At the same time, even though the authors in~\cite{vombatkere2024tiktok} provide a link to their code, the repository does not exist or is not accessible. As such, the audits cannot be easily repeated, as they can be reconstructed only based on the short description of the audits provided in the paper. In addition, without releasing the recommended items encountered during the audit, it is impossible to redo original calculations nor determine the changes on the platform, i.e., whether only the content on the platform has changed or the algorithm itself. 

Another common factor contributing to poor reproducibility is the limited description of the study design and methodology. Even though the authors often provide high-level overview of the audit, \textbf{many details are missing or are hard to judge}. In our effort to reproduce the previous audits, multiple such details were missing (e.g., how long the video should be watched in specific scenarios, or what the user characteristics are) that may have a significant effect on the overall findings. There were also inconsistencies between the paper's audit description and the released code in some cases. Without releasing the precise details about each part of the audit, the reproducibility suffers and leads to divergent findings. 

Finally, the methodological choices of the audits may also affect the reproducibility and the validity of the findings. Choosing a \textbf{setup that cannot be easily reproduced}, such as focusing the audit on time-sensitive content (such as elections in a specific country and a year), \textbf{leads to a limited reproducibility of the findings.} Similarly, we have noticed that the \textbf{setup in some of the studies may lead to incorrect findings}, such as giving stronger feedback on the content we are not trying to have personalised than on the one we are interested in (for example, watching videos of interest for a shorter duration than the remaining videos). In our study design and methodology (Section~\ref{sec:methodology}), we took these potential drawbacks into consideration and proposed such adjustments to effectively address them without diverging from the original intention of the reference studies. 

All of the audit-specific factors can be easily prevented by following the reproducibility practices in the machine learning domain. As such, there should be more focus on addressing them as fully as possible in order to limit the currently widespread poor reproducibility of the algorithmic audits.

\textbf{Platform-specific.} The platform-specific factors may have an even larger impact when it comes to poor reproducibility. First,\textbf{ the content on the platforms constantly evolves.} Even audits started a few days apart may lead to different findings, only based on the evolving popularity of the content. To mitigate this factor, more focus should be dedicated to controlling for content diversity when auditing. This is further exacerbated by different world events, such as elections, where the content may be biased by it. For example, in the reference study~\cite{boeker2022empirical}, the automated bots were instructed to focus on a sports event held that year. However, since we were reproducing such an audit a few years later, we had to change the original setup as there would be a very low probability that any videos from such an event would be recommended now.

Second, \textbf{the platform itself is constantly evolving.} For example, \textbf{new functionalities are introduced, such as livestreams or more integrated ads}, which may affect the overall scenario, what the bot sees, and how it interacts. Compared to the reference studies, we have encountered a significantly larger number of livestreams (about 1 in 20 videos) and advertisements (about 1 in 4 videos). Similarly, the HTML structure of the platform changes regularly, which often makes the resources released by previous audits unusable without further modifications. In addition, even while performing our own study, the HTML structure has changed a few times. We also observed that it is different across different countries. Such changes and differences require constant modification and limit reproducibility. Although one solution would be to rely on other approaches (e.g., computer vision approaches) that may be more robust, the audit reproducibility may still be affected. 

Finally, we have also noticed that \textbf{the platforms are starting to fight the automated bots more actively} to prevent malign coordinated behaviour (especially FIMI activities), which negatively influences algorithmic auditing. We have noticed that some of the bots (in 5 cases) were banned by the platform towards the end of the audit. In some countries (such as Italy), the banning was so prevalent and fast that it prevented us from running the audit. After the study, we performed additional checks on the accounts and observed that 15 more accounts were banned after finishing the audit study (mostly for scenarios with only implicit actions). By performing a more detailed analysis (full analysis is available as part of the released codebase), we have observed that a possible culprit may have been a problematic proxy server setup, as in some cases, the proxy server was not located in the country indicated by the service. As such, when using proxy servers, it is important to check their quality. While banning bots is a natural goal of platforms in the case of bot farms and malign coordinated behaviour, there should be an option to prevent it when bots are used for legitimate research use cases, such as algorithmic audits. 

As a result of potential (shadow)banning, \textbf{it is important to perform post-audit checks to determine its integrity.} One possibility is to request the data through GDPR (similar to~\cite{zannettou2024analyzing}) and determine whether the actions taken by the bot are recorded there. Performing such a check, we have observed that while \textbf{the watch history had an almost exact match, the explicit actions (like and follow) were missing for most of the accounts.} Similar behaviour can be observed, however, for non-bot accounts that interact solely through the webpage, while the explicit actions taken through mobile applications appear as expected. This has further negative effects on the reproducibility of the audits, as the explicit actions may not be taken into consideration by the recommender system when using the webpage and require the audits to use the mobile application (which introduces significant complications).

\section{RQ2: Effects of Personalisation Factors}
\label{sec:results_rq2}

\begin{figure*}[t]
    \centering
     \begin{subfigure}[b]{0.35\textwidth}
         \centering
         \includegraphics[width=\textwidth]{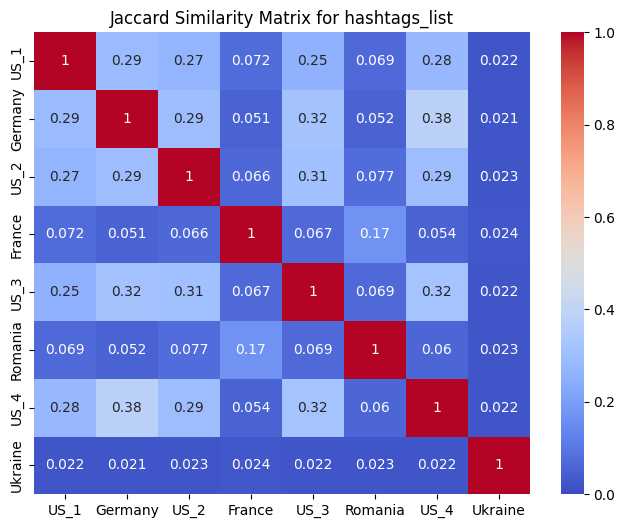}
         \caption{Jaccard similarity}
         \label{fig:location_jaccard}
     \end{subfigure}
     \begin{subfigure}[b]{0.35\textwidth}
         \centering
         \includegraphics[width=\textwidth]{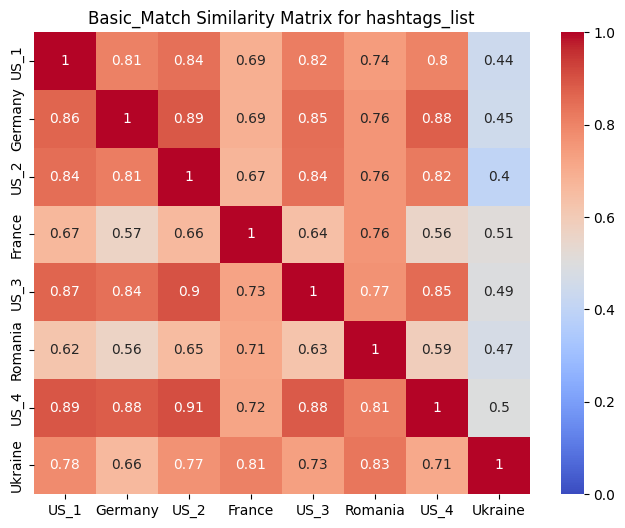}
         \caption{Basic match similarity}
         \label{fig:location_basic_match}
     \end{subfigure}
    \caption{The similarity in feeds of control and personalised users for the \textit{location} personalisation factor.}
    \label{fig:location_similarity}
    \Description{Heatmaps showing the similarity in feeds for different control and personalised users for the location personalisation factors. Overall, the similarity between control users is significantly higher than between personalised users or personalised and control users.}
\end{figure*}

\textbf{Controlling for noise.} 
To estimate how many unique videos we can expect to encounter during the algorithmic audit, we use the hashtag similarity between 2 concurrently running bots (using scenario [S0]). The average Jaccard similarity between the video feeds of these users is only $10\%$, with the similarity decreasing as more and more videos are watched. At the same time, the similarity of videos for a single user stays around $20\%$ for the whole duration of the audit. Similar results are observed when comparing control users from all scenarios that share the same setup. The average similarity in video feeds between these control users is $11\%$, with similarity as low as $2\%$ and as high as $28\%$.

As such, we observe a significant increase in the number of unique videos in comparison to the reference study by~\citet{boeker2022empirical}, where the average similarity in feeds amounted to $35\%$. This suggests that the overall diversity in the recommended videos increased during the last 3 years significantly. In addition, the difference in similarity in videos of the same user and separate users that perform the exact same actions may point to a strong exploration shift in TikTok's recommender system.

\textbf{Personalisation Factor: Location.}
To explore the impact of the location personalisation factor, we use both of the defined hashtag similarity measures. The results are presented in Figure~\ref{fig:location_similarity}.

Overall, we observe a strong effect of \textit{location} personalisation factor on the recommendation. The Jaccard similarity between all the control users (all based in the USA and performing the same actions) is between $25-38\%$. On the other hand, the similarities between the personalised users themselves or when compared against the control are an order of magnitude lower. The only exception is the user from Germany, who observed videos with hashtags similar to control users. One reason for this higher similarity may be the lower number of videos, as the user was flagged by TikTok as suspicious, and we could not perform all 4 repeats.

When using the less strict basic match similarity, we observe a lower effect of the \textit{location}  factor, with the differences between the users being much less prominent, although still apparent. We still observe high similarity (around $80-90\%$) between the control users. However, the similarity between the personalised users themselves and personalised and control users is quite high as well, around $65\%$. The only exception is the user from Ukraine, which may be caused by the use of non-Latin symbols in the hashtags.

As such, we can conclude that the \textit{location} personalisation factor has a strong impact on the TikTok recommendation system. In addition, we can see that the choice of evaluation metric (or its strictness) may skew the overall results and potentially even the findings. This observation complements the findings by~\citet{mousavi2024auditing}, who found that the hashtag similarity metrics may be a good proxy for content similarity. We argue that researchers should be careful when using these metrics, due to the potential for skewing results.

\textbf{Comparison of implicit and explicit personalisation factors.}
We first use the video popularity metric to compare the impact of implicit (\textit{watch}) and explicit (\textit{like}, \textit{follow}) personalisation factors. Table~\ref{tab:popularity} presents the results separately for scenarios that choose the videos to interact at random and those that use predefined interests.

\begin{table}[t]
    \caption{The percentual increase/decrease in popularity of the last 125 videos compared to first 125 videos.}
    \label{tab:popularity}
    
    \centering
    \begin{tabular}{@{}lcc@{}}
        \toprule
        \multirow{2}{*}{\textbf{Activity}} & \multicolumn{2}{c}{Popularity diff. in first and last 125 videos} \\
                                 & \textbf{Control} & \textbf{Personalised} \\
        \midrule
        Random Explicit & -18.30\% & -67.15\% \\
        Predefined Explicit & -66.15\% & -63.80\% \\
        Random Implicit & -37.60\% & -2.11\% \\
        Predefined Implicit & -64.35\% & -69.73\% \\
        \bottomrule
    \end{tabular}
\end{table} 
In all cases, we observe a decrease in popularity for control and personalised users. However, we do not observe a consistently larger decrease in popularity for the personalised user. Only in the case of the random explicit feedback is the decrease in popularity of videos for the personalised user significantly larger than the control user ($-67\%$ and $-18\%$). In other cases, the decrease is even lower than for the control users ($-2\%$ for the personalised users and $-37\%$ for the control users in case of random implicit feedback) or similar for both users (around $66\%$). These results may be due to the choice of the hashtags and creators used for personalisation that have been adopted from the reference study~\cite{boeker2022empirical}. As we use a rather general set of popular hashtags and creators, the video popularity may not decrease even in the case of personalisation.

To determine whether this is due to the chosen metric or the choice of personalisation, we use the hashtag basic match similarity metric. The results are available in Table~\ref{tab:hashtag-similarity-comparison}. We observe that the video similarity in the feeds of the control users remains stable, while it changes steadily for the personalised users. This indicates that the personalisation factors have an impact on the recommendation. However, for the personalised user, we observe a decrease in similarity as compared to the control user only in the case of the watch action ($-2.06\%$ for watching random videos and $-1.35\%$ for watching videos with predefined hashtags). For the other actions, we observe that the similarity stays approximately the same or even increases (especially for random liking). Although for the \textit{follow} action, we observe an increase in similarity between control and personalised user feeds, the intra-similarity of the personalised user video feed is decreasing, which indicates personalisation. As such, we can conclude that \textit{watch} is the strongest personalisation factor, followed by the \textit{follow} action. The \textit{like} action provides a personalisation signal only when defining user interests. For random \textit{like} actions, we instead observe a strong exploration.

\begin{table}[t]
\centering
\caption{The difference in hashtag basic match similarity between the first 125 and last 125 videos. The difference shows the inter and intra-change in video feed similarity for the Control (C) and Personalised (P) users.}
\label{tab:hashtag-similarity-comparison}
\begin{tabular}{@{}lccc@{}}
\toprule
\multirow{2}{*}{Activity} & \multicolumn{3}{c}{Similarity diff. in first and last 125 videos} \\
                          & C vs. P                 & C vs. C                 & P vs. P                 \\ \midrule
Follow                    & +1.50\%                 & +0.05\%                 & -2.25\%                       \\
Random Like               & +9.06\%                 & -0.33\%                 & -0.00\%                         \\
Predefined Like           & +0.04\%                 & +0.04\%                 & -0.86\%                         \\
Random Watch              & -2.06\%                 & +0.06\%                 & -1.58\%                         \\
Predefined Watch          & -1.35\%                 & +0.40\%                 & -0.28\%                       \\ \bottomrule
\end{tabular}
\end{table}

These findings are in contrast to the references studies by~\citet{boeker2022empirical} and~\citet{vombatkere2024tiktok}, where the authors found the \textit{follow} action to be most impactful and \textit{watch} only providing weaker personalisation on the level of \textit{like}. Instead, we find that just watching videos provides a stronger signal than following creators. This may be caused by a stronger exploration aspect in the explicit actions. Our findings highlight and reinforce the observation that the potential for platforms to lead users to more niche and potentially extreme content is quite strong, as the users can be led there without any explicit actions -- just by watching some videos longer.

\begin{figure}
    \centering
    \includegraphics[width=1\linewidth]{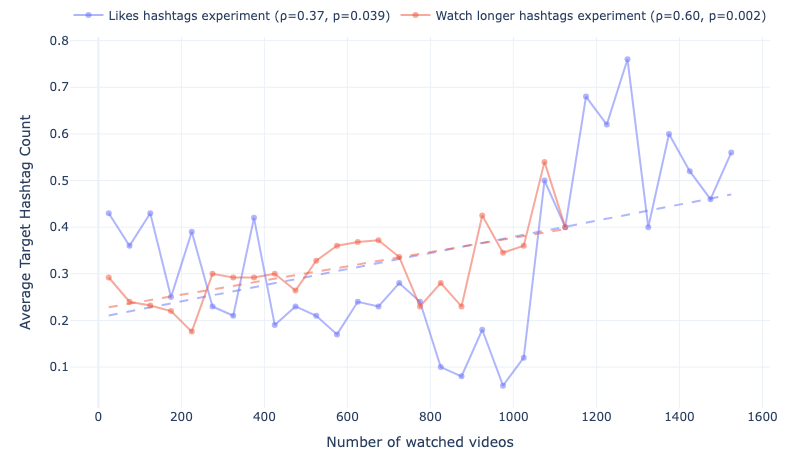}
    \caption{Comparison between the \textit{like} and \textit{watch} actions using the percentage of videos that contain predefined user interests. We can observe a steady increase for the \textit{watch} action and a strong exploration aspect for the \textit{like} action at the start, followed by strong exploitation.}
    \label{fig:like-watch-difference}
    \Description{Figure showing a comparison between like and watch actions using the percentage of videos that contain predefined user interests. For each line a trend line is also plotted. For the watch action, the line steadily increases. For the like action, the line steadily decreases until 1000 videos are watched and afterwards rapidly increases, overcoming even the line from watch action.}
\end{figure}

By looking only at scenarios with predefined user interests using hashtags ([S9], [S10], [S11] and [S13]), we can draw more nuanced findings. Instead of looking at the hashtag similarity, we count the percentage of videos containing any predefined hashtags (or their substrings). The average hashtag ratio for the \textit{watch} and \textit{like} scenarios are provided in Figure~\ref{fig:like-watch-difference}. In the case of \textit{watch} action, the ratio of videos containing the personalisation hashtags steadily increases. On the other hand, for the \textit{like} action, we observe that the ratio decreases up to a certain point (around 1000 videos), after which it increases significantly and even overtakes the ratio observed for \textit{watch} action. This suggests that in case of explicit action, the TikTok recommender system prefers exploration at the start (trying to push the users towards larger set of topics), but switches to strong exploitation after interacting with enough videos (around 1000). However, it is important to note that the comparison between explicit and implicit action may be biased by the recommender not taking explicit actions into consideration when browsing through the webpage interface (as no such action is listed in the GDPR data).

\textbf{Ratio of videos of interests.} To compare with the exploration vs. exploitation findings from the reference study by~\citet{vombatkere2024tiktok}, we use the scenarios with predefined interests ([S9]-[S11], [S13], [S14]). The ratio of videos containing at least one of the hashtags is around $36\%$ (regardless of the \textit{watch} or \textit{like} action), while the ratio of videos containing at least one of the creators is $6\%$. This is similar to findings of~\citet{vombatkere2024tiktok} that the videos of interest are presented in $30-50\%$ of cases. However, as we observe a lower end of the interval and a significantly lower ratio of videos for the \textit{follow} action, the recommender system may have shifted to a stronger exploration phase for the users at the start. Determining whether there is a change in the effect of different personalisation factors would require longitudinal audits.

\begin{figure}
    \centering
    \includegraphics[width=1\linewidth]{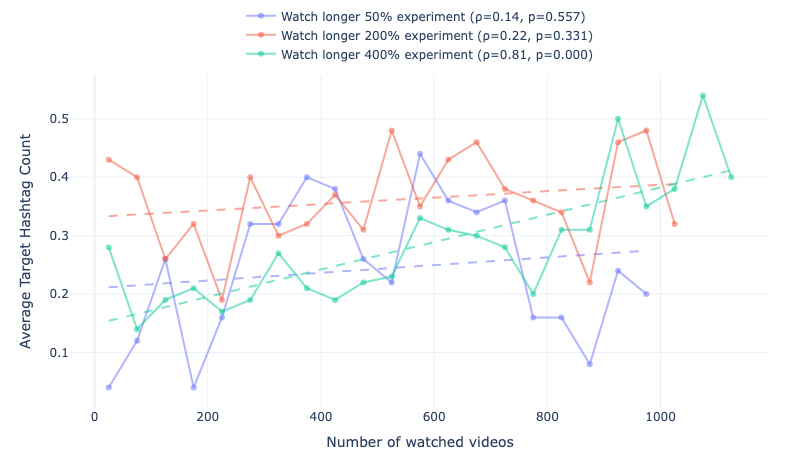}
    \caption{Comparison between the effects of different watch durations using the percentage of videos that contain predefined user interests. Watching videos for longer provides a stronger personalisation effect.}
    \label{fig:watch-duration-effect}
    \Description{Figure showing three lines that represent the percentage of videos that contain predefined user interests for different watch durations. For each line a trend line is also plotted. The trend line for the scenario where the video is watched for 400\% of its duration is the steepest.}
\end{figure}

\textbf{Effect of different watch duration.} Finally, we analyse how the different watch times affect the level of personalisation. We use the scenarios with predefined interests ([S9]-[S11]) and report the ratio of videos that contain at least one hashtag (or its substring) from our predefined list of hashtags. The results are presented in Figure~\ref{fig:watch-duration-effect}. Watching less than 100\% of the video does not seem to have a strong impact on the personalisation. Curiously, watching the videos for 200\% of their overall time has only slightly increased impact on personalisation (similar to watching 100\% of the duration). However, watching for a significantly longer duration (400\% in our case) shows a strong impact on the personalisation and the increase in the ratio of hashtags. The same behaviour can be observed when watching videos at random ([S5]-[S8]) and comparing the hashtag similarity (e.g., watching for longer leads to a stronger decrease in video similarity between the start and end of the scenario, and increased similarity between videos towards the last phase).

Again, these findings are in contrast to the reference studies that found that increasing the watch time above 75\% does not lead to a significant increase in the impact on personalisation. The difference in findings may be explained by a change in the TikTok recommender system but also by methodological issues in the reference studies (e.g., watching videos of interest for a shorter proportion of their duration than the irrelevant videos).

%% file: 5-ethical-considerations.tex
\section{Ethical Considerations and Limitations}
\label{sec:ethics}

There are several ethical and legal considerations in the study of this type. First and foremost, the execution of sockpuppeting audits requires creating automated bots and using them for data collection, which is a potential violation of the terms of services of the social media platforms. However, this breach of ToS is permitted by Article 40 (12) of the EU Act on Digital Services (DSA) if the research concerns systemic risks. Understanding the factors that influence the spread of harmful content (in our case personalization factors embedded in the social media recommender systems) can be viewed as a systemic risk, as foreseen by Recital 83 of the DSA. Second, the interaction of the bots with the content on the platform may impact the platform and society (e.g., increasing view or like count). However, we minimise the number of bots that we run, focus on popular videos where the added interactions have a minuscule impact, and after completing the study, we reverse any action that is reversible. When it comes to data, we collect only publicly available metadata, perform no annotation and use no AI. Finally, to minimise any potential legal and ethical issues, we directly involve legal and ethics experts as part of this project.

\textbf{Limitations.} As the most significant limitation, we consider the use of hashtag-based metrics, which are, however, determined by the replication character of this work. As we have shown, changing what metric is used and how strict we are when calculating the hashtag similarity has a strong effect on the findings. As such, the findings in this paper may be skewed by the used metrics. In addition, for defining the user interests, we use a set of very generic hashtags, which may affect the findings (and the used metrics). Finally, for some of the scenarios, the TikTok platform flagged our bots as suspicious, which prevented us from collecting all videos. As such, the findings from some of the scenarios may be potentially biased as they are drawn from a lower number of collected videos.

%% file: 6-conclusion.tex
\section{Discussion and Conclusion}
\label{sec:conclusion}

In this work, we focus on reproducing the algorithmic audit in order to determine the reproducibility and generalisability of the results and findings of these audits. We specifically focused on the popular social media platform, TikTok, and tried to reproduce the results of~\citet{boeker2022empirical},~\citet{vombatkere2024tiktok} and partly also~\citet{mousavi2024auditing}. Based on our experiments and reproducibility efforts, we observe two main challenges for the algorithmic audits of this kind: \textbf{poor reproducibility} and \textbf{short-term validity of the findings}. 

\textbf{Poor Reproducibility.} We identify multiple factors that negatively affect the audit reproducibility, stemming both from the platforms themselves, but also from the audit methodologies. The constantly evolving platform content, presence in multiple countries with different policies on AI or recommender systems, evolving platforms due to the added functionality or due to the policy changes, or the active fight of the platforms against automated bots -- all these factors contribute to the poor reproducibility. Even during the short time, when we were working on the audit reproducibility, we had to modify the underlying audit code to adapt to these changes. The poor reproducibility is further exacerbated by the limited resources released by the previous studies, limited description of their study design and methodology, or different methodological choices. As such, the algorithmic audits in the current state cannot be easily repeated and need to be constantly adjusted, which limits their usability and requires extensive work. 

\textbf{Short-term Validity of Findings.} Reimplementing and rerunning the reference studies, we find a significant change in the overall findings. Compared to these studies, we have found that the impact of the \textit{watch} action is significantly stronger, which is further increased as the video is watched for longer or multiple times. At the same time, we observe a shift to a stronger exploration component for the \textit{like} and \textit{follow} action, which reduces their personalisation impact in the first set of around 1000 videos. However, we also observe that some explicit actions may not be taken into consideration by the recommender algorithm when using TikTok's web interface, as these actions do not appear as part of the requested GDPR data. Most significantly, the findings are strongly dependent on the methodological setup, where just by changing the evaluation metric or how strict we are when computing similarity may lead to completely different findings. A similar effect can be observed based on small variations in the bot simulation.

Following the findings of our paper, we advocate for reproducible, longitudinal, multiplatform audits with more authentic user simulation that can more faithfully discern the changes in findings.